**Title Page:**

**Title:** Multiomics-based Outcome Prediction for the Treatment of Brain Metastases with Personalized Ultra-fractionated Stereotactic Adaptive Radiotherapy

**Running title:** Multiomics for PULSAR Outcome Prediction

**Authors:** Haozhao Zhang, MS[1]*, Michael Dohopolski, MD[1]*, Strahinja Stojadinovic, PhD[1], Luiza Giuliani[1], Soummitra Anand[1], Heejung Kim, PhD[1], Arnold Pompos, PhD[1], Andrew Godley, PhD[1], Steve Jiang, PhD[1], Tu Dan, PhD[1], Zabi Wardak, MD[1], Robert Timmerman, MD[1], Hao Peng, PhD[1]#

* These authors contributed equally to this work.

# Corresponding author

**The author's institutional affiliations**:

[1] Department of Radiation Oncology, The University of Texas Southwestern Medical Center, Dallas, TX 75390, United States of America

**Contact information:**

E-mail: Hao.Peng@utsouthwestern.edu

## Abstract:

**Background:** Brain metastases (BM) are a prevalent complication in patients with systemic cancer. Conventional treatments have limited efficacy in addressing multifocal microscopic disease and are associated with significant neurotoxicity, particularly in larger tumors. Personalized ultra-fractionated stereotactic adaptive radiotherapy (PULSAR) has emerged as a promising approach, delivering high-dose radiation in extended intervals to allow for greater normal tissue recovery and leverage adaptive therapy to enable precise, personalized treatment adjustments based on individual responses. However, current PULSAR plan adaptations rely primarily on physicians' experiences and assessments of tumor size changes observed in intra-treatment MRI scans, which introduces significant uncertainty, potentially leading to over-treatment or under-treatment, thus limiting the full potential of PULSAR. This underscores the need for an objective and robust method to enhance treatment outcome prediction then assisting with decision-making in PULSAR treatment.

**Purpose:** To develop a data-driven multiomics approach integrating radiomics, dosiomics, and delta features to predict treatment response at an earlier stage (intra-treatment) for BM patients treated with PULSAR.

**Methods:** A retrospective study encompassing 39 BM patients with 69 lesions treated with PULSAR was undertaken. Radiomics, dosiomics, and delta features were extracted from both pretreatment and intra-treatment MRI scans alongside dose distributions. Six individual models, alongside an ensemble feature selection (EFS) model, were constructed utilizing support vector machines (SVM) and evaluated via stratified 5-fold cross-validation. The classification task aimed to distinguish between two lesion groups, depending on whether they exhibited a volume reduction of more than 20% at follow-up. Performance metrics were assessed, including sensitivity, specificity, accuracy, precision, F1 score, and the area under the ROC (Receiver Operating Characteristic) curve (AUC). Various scenarios of feature extraction and ensemble feature selection were explored to bolster model robustness and mitigate overfitting.

**Results:** The EFS model integrates crucial features from pre-treatment radiomics, pre-treatment dosiomics, intra-treatment radiomics, and delta-radiomics. It surpasses six individual models, achieving an AUC of 0.979, accuracy of 0.917, and F1 score of 0.821. Among the top 9 features of the EFS model, six features come from post-wavelet transformation, and three from original images. The discrete wavelet transform decomposes volumetric images into multi-resolution components, providing a more comprehensive characterization of the underlying structures.

**Conclusions:** Our study demonstrated the feasibility of employing a data-driven multiomics approach to predict tumor volume changes in BM patients receiving PULSAR treatment. The EFS model demonstrates enhanced performance compared with six individual models, emphasizing the importance of integrating both pretreatment and intra-treatment data. This application of multiomics alongside SVM classification for intra-treatment decision support in PULSAR holds promise for optimizing BM management, potentially mitigating risks associated with under- or over-treatment.

## 1. Introduction

Brain metastases (BM) represent the most common form of intracranial tumors, afflicting 20 to 40% of patients with systemic cancer.[1-4] Surgical intervention and radiotherapy have emerged as principal treatment modalities, as many chemotherapies do not cross the blood brain barrier.[5] Surgical resection offers immediate mass effect relief but is often insufficient for eradicating multifocal microscopic disease. Stereotactic radiosurgery (SRS) utilizes highly focused radiation; however, its efficacy is inversely correlated with tumor size and poses a heightened risk of neurotoxicity.[4-6] The field has transitioned to fractionated stereotactic radiotherapy (fSRT) and staged stereotactic radiosurgery (SSRS) for treating larger BM, which improves local control while minimizing adverse radiation effects.[7-10] Nevertheless, the determination of optimal treatment intervals, particularly for aggressive tumors, still remains an ongoing subject of investigation.[4]

At our institution, we have adopted a similar approach to SSRS called personalized ultra-fractionated stereotactic adaptive radiotherapy (PULSAR), where we deliver high dose radiation in two to four-week intervals. Not all cancers are the same, and each patient responds differently to treatment. Longer intervals (weeks or months) allow for more normal tissue recovery after injury, while concurrently providing time for the tumor and tumor microenvironment (TME) to undergo dramatic changes.[11,12] This consequently allows for much more meaningful adaptation that considers the evolving characteristics of the tumor, such as size, shape, and biomarker expression changes. Another point of interest is the potential synergy between PULSAR and checkpoint blockade inhibitor immunotherapy.[11]

To maximize the potential benefits of PULSAR, a crucial aspect involves decision-making, either before or during treatment (**Figure 1**). The current procedure largely relies on the physician's expertise and experience. For instance, the treatment plan may be adjusted depending on the change in gross tumor volume (GTV) from the intra-treatment MRI scan (e.g. 20% change).[13-15] However, such a simple metric may not correlate well with the ultimate treatment outcome. It is thus desirable to develop a more objective and quantitative method to assist with decision-making in PULSAR treatment, minimizing the risks of under-treatment or over-treatment.[16,17] To achieve this objective, our study employed a data-driven multiomics approach for predicting treatment outcome in terms of tumor volume change. Unlike most previous outcome prediction studies that rely solely on pretreatment images, our investigation aimed to utilize radiomics and dosiomics features extracted at different time points (i.e. in a delta mode) to enhance predictive power, due to the distinctive features of PULSAR.[18-24] We hypothesize that by leveraging additional information available one month after the initial treatment, the predictive model would be able to achieve improved accuracy and assist physicians with decision-making more efficiently.

## 2. Methods

### 2.1 PULSAR Treatment, Study Population and Data Acquisition

**Figure 1** provides a workflow comparison between fSRT and PULSAR for patients with BM undergoing Gamma Knife radiosurgery. In the PULSAR treatment process, the patient initially undergoes a pretreatment

MRI scan, followed by the first treatment course comprising three fractions/pulses (5 to 6 Gy per fraction/pulse) with a two-day interval between fractions. After two to four weeks, the second course is delivered according to the intra-treatment MRI scans, with adjustments for changes in tumor volume and/or presence of vasogenic edema. Mostly commonly, the plan is adjusted to a smaller target or unchanged target. Uncommonly, treatment is halted if a target is no longer visible, or the treatment is adjusted for a target of increasing size through either a dose boost or surgical intervention.

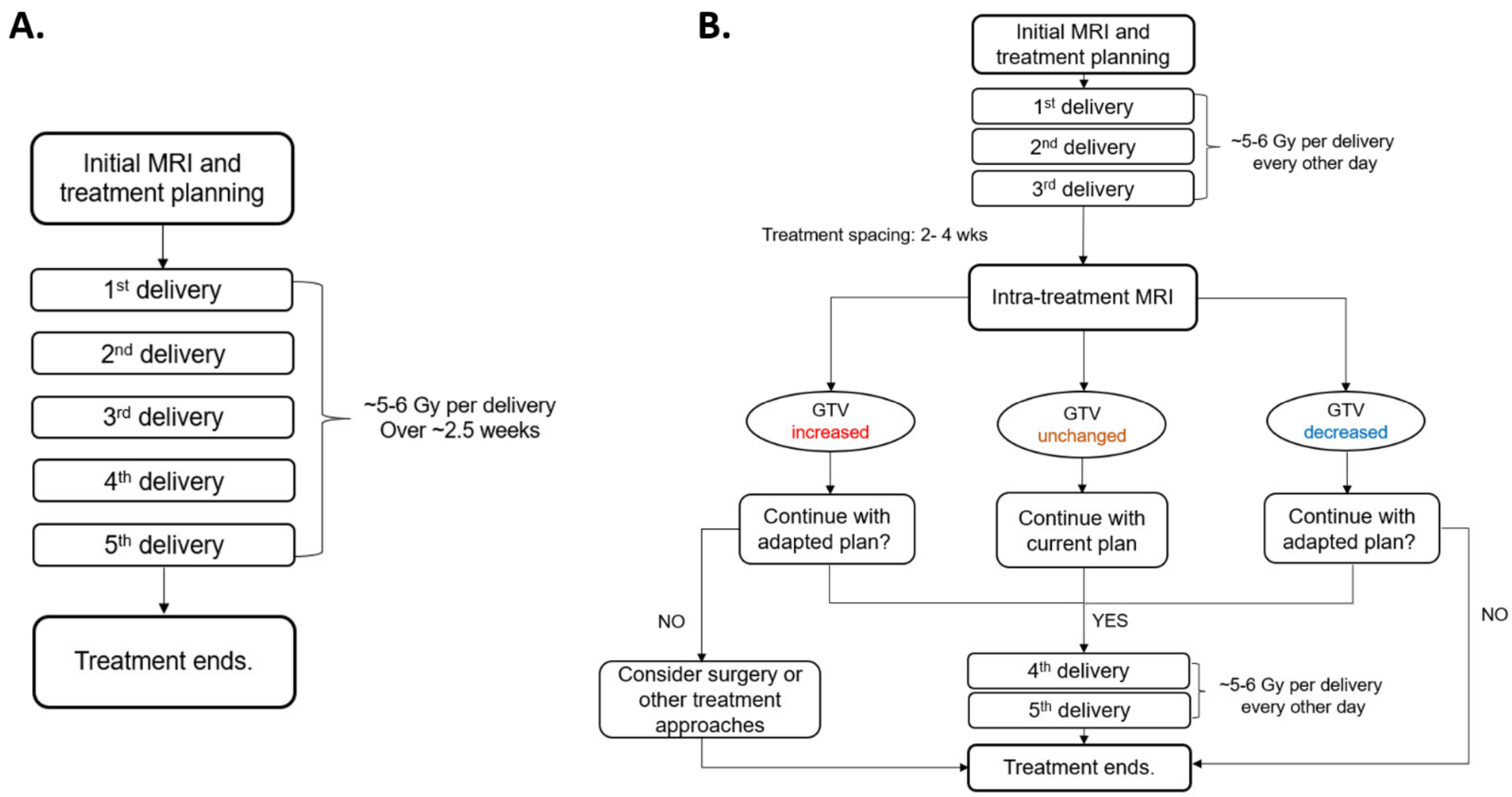

**Figure 1**. Comparison of workflow between (A) fractionated stereotactic radiotherapy (fSRT) and (B) personalized ultra-fractionated stereotactic adaptive radiotherapy (PULSAR). PULSAR includes an intra-treatment MRI assessment to evaluate the change of GTV (increased, unchanged, or decreased), enabling more personalized treatment and timely adjustment.

Our retrospective study involved the examination of 39 BM patients who underwent PULSAR Gamma Knife treatment at UTSW. The cohort comprised 69 lesions treated from November 1, 2021, to May 1, 2023, with both single and multiple BM. **Table 1** summarizes detailed demographic and clinical profiles, including age, gender, lesion number, and treatment specifics. Comprehensive initial and intra-treatment data were collected from patients undergoing PULSAR, encompassing MRI images, 3D dose maps (RTdose), and radiotherapy contour structure files (RTstructure). All collected images were acquired using axial (AX) MRI sequences with T1-weighted enhancement, ensuring a consistent and standardized basis for radiomic analysis. This approach minimized variations and potential discrepancies that could arise from different imaging modalities or sequences, thereby ensuring the accuracy of subsequent delta-omics calculations in the PULSAR cohort. Additionally, tumor volumes in follow-up MRI images were evaluated one to three months after PULSAR treatment. Treated BM were manually contoured on these follow-up scans using the Research Velocity platform. A thorough comparison of pre-, intra-, and post-treatment images for each lesion was conducted by a research assistant and subsequently reviewed by an experienced physician.

**Table 1. Patient Demographics and Lesion Characteristics in PULSAR Treatment.**

| Characteristics | Value |
|---|---|
| **Patient Information** | |
| Age (range) | 61 (28-84) |
| Total Patients (Male: Female) | 39 (14:25) |
| **Lesion Information** | |
| Single brain metastases | 26 |
| Multiple brain metastases | 43 |
| Total Lesions (Decreased: Non-Decreased) * | 69 (55:14) |
| **Lesion size** | |
| Initial Tumor Volume ($mm^3$) | 3666.3 (14.5-37607.6) |
| 3D Diameter (mm) | 25.5 (3.7-79.8) |
| **Delivered Dose and Fractionation by Lesion Count (Gy, Fx)** | |
| 30 Gy, 5Fx | 57 |
| 27.5 Gy, 5Fx | 2 |
| 25 Gy, 5Fx | 9 |
| 24 Gy, 4Fx | 1 |

* Threshold for change defined as 20% GTV reduction.

## 2.2 Data Processing and Multiomics Feature Extraction

Pre-processing steps were executed to enhance the reliability and consistency of radiomics analysis, including image resolution adjustment and co-registration (MRI image, RTstructure, RTdose). 3D MRI images, 3D dose maps, and 3D GTVs were co-registered. Subsequently, they were re-sampled to standardize voxel sizes to a uniform 1×1×1 $mm^3$, ensuring accuracy and consistency in feature extraction. Feature extraction was divided into two parts: 1) Radiomics features: 1st radiomics features (from pre-treatment MRI) and 2nd radiomics features (from intra-treatment MRI) were extracted using PyRadiomics, with each scenario yielding 851 features; 2) Dosiomics features: 3D dose maps were treated in the same manner as MRI images, for the extraction of 1st dosiomics features and 2nd dosiomics features using PyRadiomics, also yielding 851 features per scenario.[25] These features encompassed various categories, including intensity, shape, texture, and wavelet filters, providing a comprehensive characterization of the tumor's imaging properties and dose characteristics. After extraction, both radiomics and dosiomics features underwent z-score normalization. Furthermore, delta-radiomics and delta-dosiomics features were computed by subtracting the first radiomics/dosiomics features from the second radiomics/dosiomics features after z-score normalization, quantifying the change in tumor characteristics during treatment and providing insights into tumor response.

## 2.3 Multi-Level Feature Selection (MLF)

Given the complexity and high dimensionality of the radiomics and dosiomics feature sets (with 851 features extracted per scenario), a robust MLF strategy was implemented in three sequential steps. The first step was the filtering method, which gauges the statistical dependence between each feature and the volume change based on Spearman's correlation coefficients. The top 50% of features with the highest absolute correlation values were retained for further analysis. The second step was the embedded method using LASSO (Least Absolute Shrinkage and Selection Operator) regression, which introduced an L1 regularization penalty to the loss function for reducing the number of features.[26] The third step was recursive feature elimination (RFE), combined with a Support Vector Machine (SVM) classifier, to rank features based on their importance and iteratively remove the least important ones.[27,28] To minimize potential bias and overfitting in feature selection, given the small size of the dataset, we performed MLF with 5-fold stratified cross-validation (50 iterations). During each iteration, feature selection was conducted solely on the training folds, ensuring the test fold remained unseen. The top 9 features were recorded for each iteration, and the 9 most frequently occurring features across all iterations were identified.

### 2.4 Predictive Model and Performance Evaluation

Our model predicts tumor volume change, specifically if a lesion is expected to show a volume reduction of ≥ 20% at the time point of follow-up. Several studies suggest a significant correlation between a reduction in volume of 20% or greater and the improvement of neurological signs and symptoms, reproducible when interpreted by different clinicians.[13-15] We adopted the same criterion to categorize lesions and formulated it as a classification problem. Lesions with a follow-up tumor volume equal to or greater than 80% of their initial volume were classified as 'non-decreased volume' (*Group A*), while those with a reduction below this threshold were categorized as 'decreased volume' (*Group B*). Equipped with features after MLF, we trained six individual models. The individual models consisted of features extracted from 1) pretreatment images ("1st radiomics"), 2) intra-treatment images ("2nd radiomics"), 3) the change between pretreatment and intra-treatment radiomics ("delta-radiomics"), 4) pretreatment dose distributions ("1st dosiomics"), 5) intra-treatment dose distributions ("2nd dosiomics"), and 6) the change between pretreatment and intra-treatment dosiomics ("delta-dosiomics"). Each model, utilizing the nine most frequently discriminative features identified after the MLF described in section 2.3, was fitted using a support vector machine (SVM) for classification. Optimal SVM parameters were identified through grid search and 5-fold cross-validation. Due to the small size of the dataset (as well as being imbalanced), each model was trained and validated via stratified 5-fold cross-validation, maintaining the ratio between the two groups constant in each fold. To assess the robustness and stability of each model, the 5-fold cross-validation procedure was repeated 50 times with different random seeds. Performance metrics, including sensitivity, specificity, accuracy, precision, F1 score, and AUC, were calculated for each repetition. Aggregated metrics across the 50 repetitions, such as mean values, standard deviations, and 95% confidence intervals, were used to evaluate overall model performance and its fluctuations. Furthermore, 3-fold cross-validation was conducted in the same manner to check for consistency (referring to **Figure S1** in supplemental materials).

### 2.5 Ensemble Feature Selection (EFS) and Model Comparison

To further explore the potential of the synergistic power of multiomics and fully leverage the available information at the intra-treatment time point, which includes pretreatment image, intra-treatment image, and pretreatment dose, an EFS strategy was implemented to identify the most prominent features from four scenarios: 1st radiomics, 2nd radiomics, delta-radiomics, and 1st dosiomics, utilizing the same MLF strategy described in section 2.3. Subsequently, similar to the approach used for developing individual models, an ensemble model was developed also employing SVM (hereafter referred to as the EFS model). For the EFS model, the same quantitative analysis was conducted and subsequently compared with the six individual models. In total, seven SVM-based models were trained and evaluated. To interpret the output of the SVM models probabilistically, the Platt scaling was used to evaluate the prediction probability of each lesion.[29]

## 3. Results

### 3.1 Treatment response

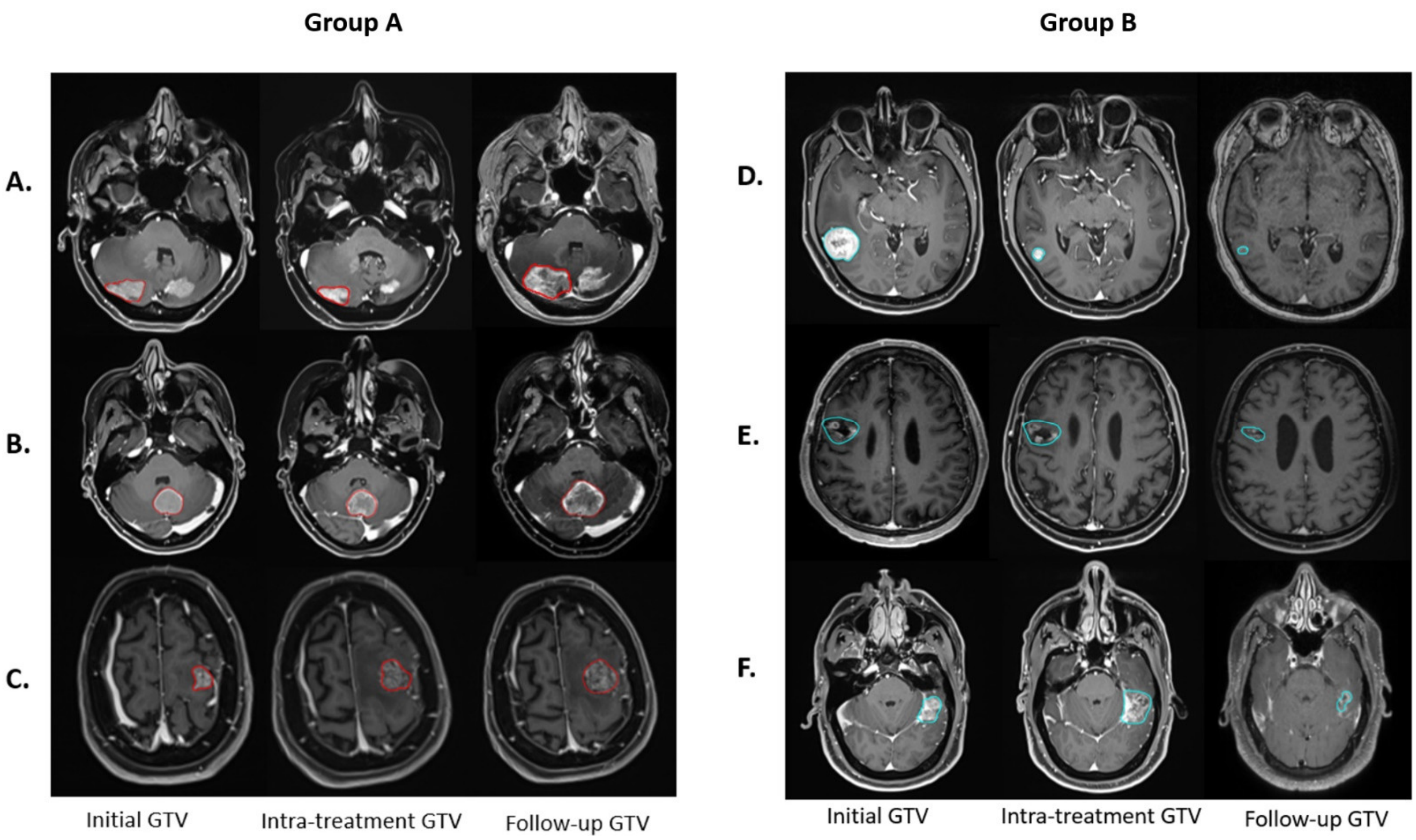

**Figure 2.** Six lesions illustrate the temporal evolution of GTV at various treatment stages. In Group A (lesions with red contours), three lesions exhibit non-decreased GTV at follow-up compared to the initial, but they display different GTV changes at the intra-treatment time point, with intra-treatment assessments of (A) decreased, (B) unchanged, and (C) increased GTV. In contrast, Group B (lesions with blue contours) depicts three lesions with a decreased GTV at follow-up compared to the initial, with intra-treatment variations categorized as (D) decreased, (E) unchanged, and (F) increased GTV.

Volumetric analysis elucidates a multifaceted response to PULSAR treatment. **Figure 2** shows six representative lesions, each demonstrating the change of tumor volume at three time points. The dynamics suggest that intra-treatment evaluations alone, categorized as increased, unchanged, or decreased, do not adequately reveal treatment outcome. To quantify GTV change, **Figure 3** maps out the volumetric dynamics of all 69 lesions at pre-treatment, intra-treatment, and follow-up time points, with the initial tumor volume normalized to 1 for each lesion. The analysis stratifies the lesions into two subsets based on a threshold of 4000 mm³: 36 large lesions (**Fig. 3A**) and 33 small lesions (**Fig. 3B**). Among the large lesions, 5 out of 36 exhibit an increase in intra-treatment volume with the highest ratio of 1.18, and 8 lesions display a ratio in the range between 0.8 and 1.0, indicating no substantial volume change. The intra-treatment volume for the small lesions has a single case showing a ratio of 1.06, and 11 lesions between 0.8 and 1.0. When examining the last column in the heatmaps (follow-up GTV), for the subset of large lesions, 6 lesions exhibit an increase (highlighted with red diamonds) with the largest ratio of 2.06, while 3 lesions achieve a moderate reduction in volume with the ratio between 0.8 and 1.0. In the smaller lesion subset, 3 lesions demonstrate minimal change with ratios between 0.8 and 1.0, and two lesions increase in size (also highlighted with red diamonds), with the largest ratio at 2.28.

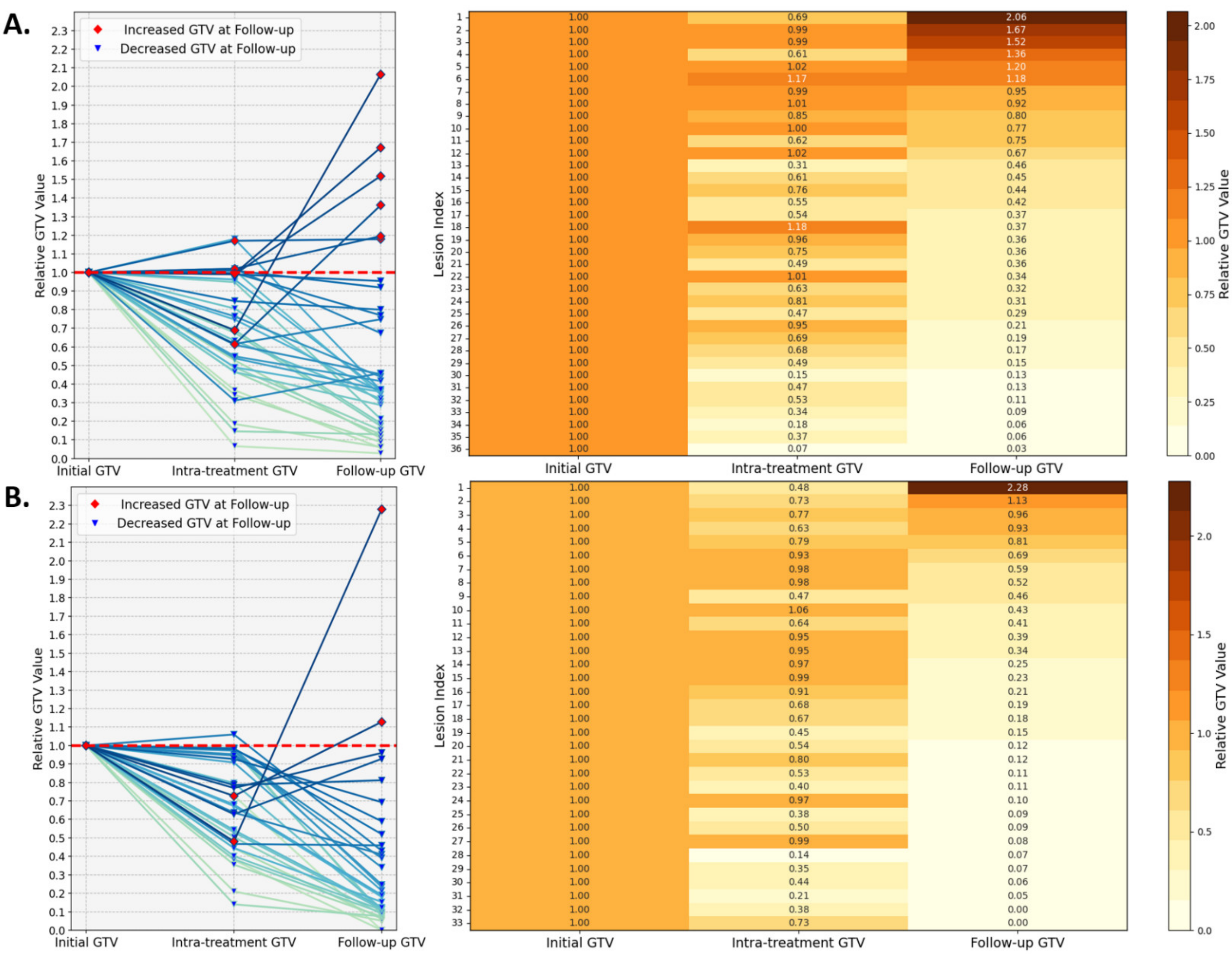

**Figure 3.** Lesion volumetric changes are examined at three time points (pre-treatment, intra-treatment, and follow-up) for two subsets: lesions smaller than 4000 mm³ (Fig. 3A) and those larger than 4000 mm³ (Fig. 3B). The volumetric dynamics of all 69 lesions are shown with the initial GTV normalized to 1 for each lesion. The line graphs on the left display the relative GTV changes for each lesion during PULSAR. The color intensity of each lesion's trajectory line reflects the relative change at follow-up compared to the initial GTV, with lighter colors representing more significant reductions after PULSAR treatment. Red diamonds indicate lesions with increased GTV at follow-up, while blue triangles represent lesions with decreased GTV at follow-up. The accompanying heatmaps on the right provide a detailed quantitative representation of these changes, with color intensity reflecting the relative increase or decrease in tumor volume.

### 3.2 Performances of Individual Models

Fourteen lesions are classified into *Group A* where there is no significant volume reduction (relative GTV ratio at follow-up ≥ 0.8), and 55 lesions into *Group B* (relative GTV ratio at follow-up < 0.8). Each model incorporates the top 9 features after MLF. The ROC curves for the six individual models are presented in **Fig. 4A** and performance metrics are summarized in **Table 2**. **Figure S2** exemplifies the correlation of multiple features for each model, with the heatmaps offering an intuitive perspective on the relationships between the selected features. **Table 3** presents the pairwise p-values calculated using Welch's t-test with Bonferroni correction for inter-model comparison, examining whether a significant difference exists between two models.

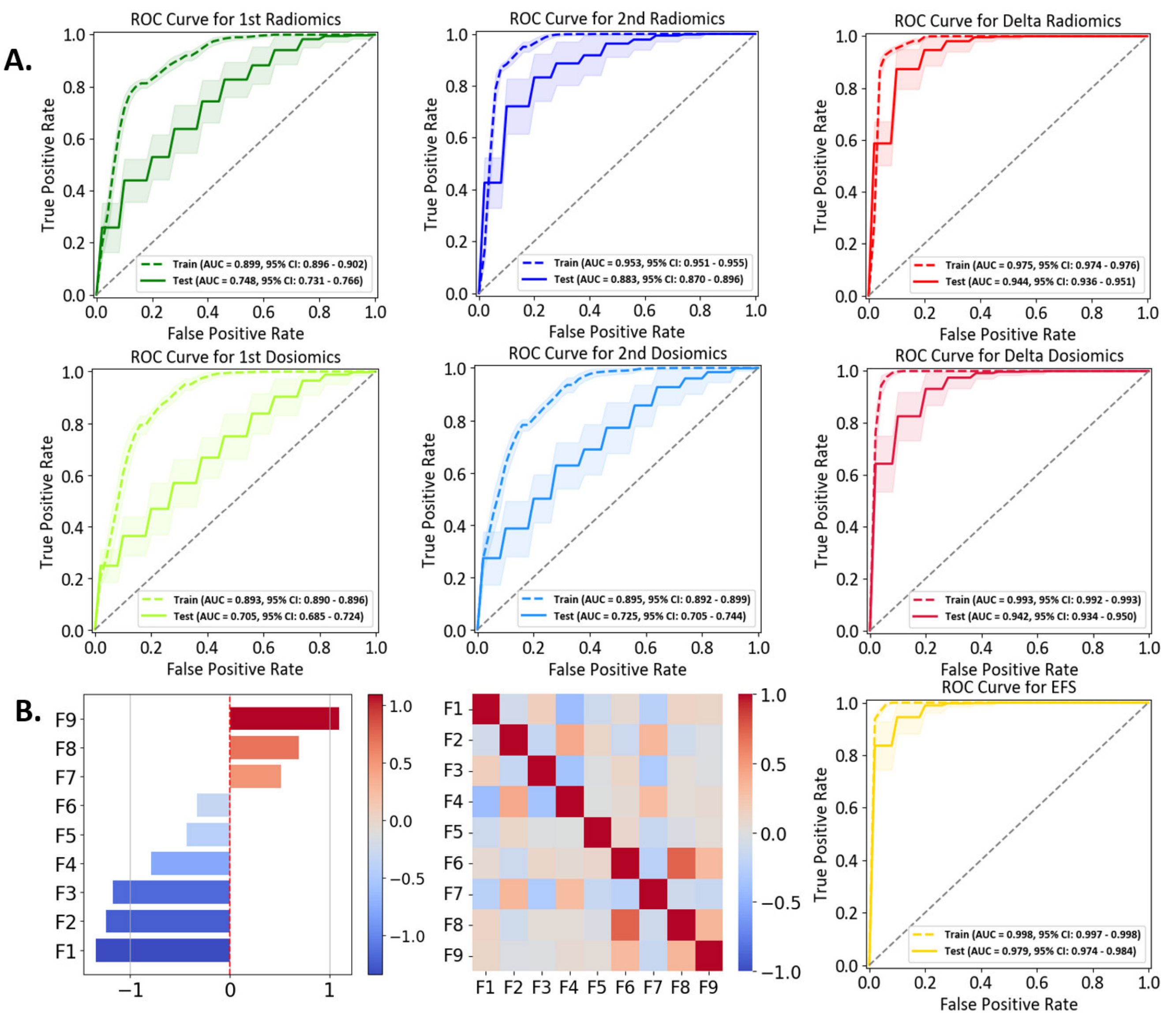

**Figure 4.** (Fig. 4A) Receiver Operating Characteristic (ROC) curves for six individual models. For each model, the plot shows the aggregated training (solid line) and test (dashed line) ROC curves generated from 50 iterations of 5-fold stratified cross-validation. The mean Area Under the Curve (AUC) with 95% confidence interval is reported for each model, respectively. (Fig. 4B) Performance evaluation of the Ensemble Feature Selection (EFS) model. The left panel displays the coefficient values of the 9 selected features in the EFS model. The middle panel presents the correlation heatmap of these 9 EFS features, demonstrating minimal inter-feature correlation. The right panel shows the ROC curve of the EFS model similar to those in (A).

From **Fig. 4A**, the 1st dosiomics model exhibits the lowest AUC of 0.705, with a low precision of 0.285. The 2nd dosiomics model exhibits the second-lowest AUC (0.725), and a low precision (0.319). Such inferior performance suggests the limitations of a model solely based on dosiomics. Two models based upon MRI image data, the 1st radiomics and the 2nd radiomics, outperform the above two counterparts, achieving AUC (0.748) / precision (0.344) and AUC (0.883) / precision (0.605), respectively. Compared to the 1st dosiomics model, the 1st radiomics model demonstrated improved performance, with significantly higher specificity, accuracy, AUC, and precision (*p*-values<0.002, **Table 3**). The 1st radiomics and 2nd dosiomics models demonstrate comparable performance, with no significant differences found between them. Among the four individual models based on single time points mentioned above, the 2nd radiomics model demonstrates the best performance.

The delta radiomics and delta dosiomics models show comparable performance, both outperforming the single time point models. The delta radiomics model excels in sensitivity (0.903 vs. 0.802), while the delta dosiomics model has higher specificity (0.836 vs. 0.872). Both models have similar accuracy (0.850 vs. 0858), AUC (0.944 vs. 0.942), precision (0.632 vs. 0.661), and F1 score (0.720 vs. 0.694). The relative change in dose distribution appears to provide additional valuable information for the classification task. By leveraging changes in features over time, the delta models better capture tumor evolution and treatment response. The ROC comparison of different models using 5-fold and 3-fold cross-validation to evaluate overall performance fluctuation, as well as an example of the confusion matrices from a single iteration of 5-fold cross-validation for the EFS model, are provided in Figures S1 and S3, respectively.

**Table 2. Performance Metrics of Different Models.**

| | 1st Radiomics | 2nd Radiomics | Delta Radiomics | 1st Dosiomics | 2nd Dosiomics | Delta Dosiomics | Ensemble feature selection (EFS) |
|---|---|---|---|---|---|---|---|
| **Sensitivity** | 0.503 ± 0.271 (95% CI: 0.469 - 0.537) | 0.755 ± 0.266 (95% CI: 0.722 - 0.789) | 0.903 ± 0.174 (95% CI: 0.882 - 0.925) | 0.559 ± 0.271 (95% CI: 0.525 - 0.592) | 0.538 ± 0.309 (95% CI: 0.499 - 0.577) | 0.802 ± 0.242 (95% CI: 0.772 - 0.832) | 0.907 ± 0.169 (95% CI: 0.886 - 0.928) |
| **Specificity** | 0.745 ± 0.128 (95% CI: 0.729 - 0.761) | 0.854 ± 0.102 (95% CI: 0.841 - 0.867) | 0.836 ± 0.117 (95% CI: 0.822 - 0.851) | 0.636 ± 0.141 (95% CI: 0.618 - 0.654) | 0.713 ± 0.129 (95% CI: 0.697 - 0.729) | 0.872 ± 0.102 (95% CI: 0.860 - 0.885) | 0.920 ± 0.085 (95% CI: 0.909 - 0.931) |
| **Accuracy** | 0.696 ± 0.107 (95% CI: 0.682 - 0.709) | 0.833 ± 0.090 (95% CI: 0.822 - 0.844) | 0.850 ± 0.094 (95% CI: 0.838 - 0.862) | 0.621 ± 0.112 (95% CI: 0.607 - 0.635) | 0.677 ± 0.111 (95% CI: 0.664 - 0.691) | 0.858 ± 0.086 (95% CI: 0.847 - 0.869) | 0.917 ± 0.072 (95% CI: 0.908 - 0.926) |
| **AUC** | 0.748 ± 0.143 (95% CI: 0.731 - 0.766) | 0.883 ± 0.107 (95% CI: 0.870 - 0.896) | 0.944 ± 0.061 (95% CI: 0.936 - 0.951) | 0.705 ± 0.157 (95% CI: 0.685 - 0.724) | 0.725 ± 0.158 (95% CI: 0.705 - 0.744) | 0.942 ± 0.062 (95% CI: 0.934 - 0.950) | 0.979 ± 0.039 (95% CI: 0.974 - 0.984) |
| **Precision** | 0.344 ± 0.197 (95% CI: 0.319 - 0.368) | 0.605 ± 0.232 (95% CI: 0.577 - 0.634) | 0.632 ± 0.197 (95% CI: 0.607 - 0.657) | 0.285 ± 0.140 (95% CI: 0.268 - 0.302) | 0.319 ± 0.182 (95% CI: 0.296 - 0.341) | 0.661 ± 0.224 (95% CI: 0.633 - 0.689) | 0.786 ± 0.201 (95% CI: 0.761 - 0.811) |
| **F1 Score** | 0.389 ± 0.193 (95% CI: 0.365 - 0.413) | 0.640 ± 0.199 (95% CI: 0.615 - 0.665) | 0.720 ± 0.151 (95% CI: 0.701 - 0.739) | 0.366 ± 0.162 (95% CI: 0.346 - 0.386) | 0.386 ± 0.205 (95% CI: 0.361 - 0.412) | 0.694 ± 0.186 (95% CI: 0.671 - 0.718) | 0.821 ± 0.147 (95% CI: 0.803 - 0.839) |

**Table 3. Pairwise Comparison of Model Performance Metrics.**

| | 1st Radiomics vs. 2nd Radiomics | 1st Radiomics vs. Delta Radiomics | 1st Radiomics vs. 1st Dosiomics | 1st Radiomics vs. 2nd Dosiomics | 1st Radiomics vs. Delta Dosiomics | 1st Radiomics vs. EFS | 2nd Radiomics vs. Delta Radiomics | 2nd Radiomics vs. 1st Dosiomics | 2nd Radiomics vs. 2nd Dosiomics | 2nd Radiomics vs. Delta Dosiomics | 2nd Radiomics vs. EFS | Delta Radiomics vs. 1st Dosiomics | Delta Radiomics vs. 2nd Dosiomics | Delta Radiomics vs. Delta Dosiomics | Delta Radiomics vs. EFS | 1st Dosiomics vs. 2nd Dosiomics | 1st Dosiomics vs. Delta Dosiomics | 1st Dosiomics vs. EFS | 2nd Dosiomics vs. Delta Dosiomics | 2nd Dosiomics vs. EFS | Delta Dosiomics vs. EFS |
|---|---|---|---|---|---|---|---|---|---|---|---|---|---|---|---|---|---|---|---|---|---|
| **Sensitivity** | < 0.002* | < 0.002* | **0.022** | **0.176** | < 0.002* | < 0.002* | < 0.002* | < 0.002* | < 0.002* | **0.041** | < 0.002* | < 0.002* | < 0.002* | < 0.002* | **0.829** | **0.428** | < 0.002* | < 0.002* | < 0.002* | < 0.002* | < 0.002* |
| **Specificity** | < 0.002* | < 0.002* | < 0.002* | **0.006** | < 0.002* | < 0.002* | **0.070** | < 0.002* | < 0.002* | **0.047** | < 0.002* | < 0.002* | < 0.002* | < 0.002* | < 0.002* | < 0.002* | < 0.002* | < 0.002* | < 0.002* | < 0.002* | < 0.002* |
| **Accuracy** | < 0.002* | < 0.002* | < 0.002* | **0.063** | < 0.002* | < 0.002* | **0.045** | < 0.002* | < 0.002* | < 0.002* | < 0.002* | < 0.002* | < 0.002* | **0.328** | < 0.002* | < 0.002* | < 0.002* | < 0.002* | < 0.002* | < 0.002* | < 0.002* |
| **AUC** | < 0.002* | < 0.002* | < 0.002* | **0.079** | < 0.002* | < 0.002* | < 0.002* | < 0.002* | < 0.002* | < 0.002* | < 0.002* | < 0.002* | < 0.002* | **0.758** | < 0.002* | **0.158** | < 0.002* | < 0.002* | < 0.002* | < 0.002* | < 0.002* |
| **Precision** | < 0.002* | < 0.002* | < 0.002* | 0.140 | < 0.002* | < 0.002* | **0.170** | < 0.002* | < 0.002* | **0.007** | < 0.002* | < 0.002* | < 0.002* | **0.127** | < 0.002* | **0.022** | < 0.002* | < 0.002* | < 0.002* | < 0.002* | < 0.002* |
| **F1 Score** | < 0.002* | < 0.002* | **0.152** | **0.882** | < 0.002* | < 0.002* | < 0.002* | < 0.002* | < 0.002* | < 0.002* | < 0.002* | < 0.002* | < 0.002* | **0.088** | < 0.002* | **0.221** | < 0.002* | < 0.002* | < 0.002* | < 0.002* | < 0.002* |

*** Significant difference is identified based on the Welch's t-test with Bonferroni correction.**

### 3.3 Ensemble Feature Selection (EFS) Model

The EFS model, developed using the strategy described in Section 2.5, comprises 9 features identified from a pool of features originating from four scenarios (1st radiomics, 2nd radiomics, delta radiomics, and 1st dosiomics). The correlation heatmap of these 9 features (shown in **Fig. 4B)** demonstrates minimal inter-feature correlation. Detailed descriptions (e.g., names, types, weights in the SVM) are summarized in **Table 4**, where the weight coefficients (in a descending order) reflect their respective influence and the 95% CIs for the z-score normalized feature values are presented for both *Group A* and *Group B*. The ROC curve of the EFS model is depicted in **Fig. 4B**, exhibiting AUC of 0.979, sensitivity (0.907), specificity (0.920), accuracy (0.917), precision (0.786), and F1 score (0.821). While the EFS model has a sensitivity comparable to that of delta radiomics model, the EFS model consistently achieves superior performance in all other metrics.

**Table 4. Summary of the Nine Selected Multiomics Features After Ensemble Feature Selection (EFS).**

| Abbreviation | Multi-omics Type | Wavelet Filtering/Original | Feature Class | Feature Name | Feature Weight Coefficient | Feature Values (Z-score) | | |
|---|---|---|---|---|---|---|---|---|
| | | | | | | Group A | Group B | P-value |
| F1 | Delta Radiomics | Original | shape | Least Axis Length | -1.34382 | -0.231 ± 0.233 | 0.059 ± 0.468 | <0.05* |
| F2 | 1st Dosiomics | Wavelet-LHL | glszm | Low Gray Level Zone Emphasis | -1.23960 | -0.58 ± 0.867 | 0.148 ± 0.994 | <0.05* |
| F3 | 2nd Radiomics | Wavelet-LHH | glcm | Correlation | -1.17479 | -0.577 ± 0.584 | 0.147 ± 1.043 | <0.05* |
| F4 | 2nd Radiomics | Original | glrlm | Long Run Low Gray Level Emphasis | -0.78866 | -0.444 ± 0.195 | 0.113 ± 1.098 | <0.05* |
| F5 | Delta Radiomics | Wavelet-HHH | gldm | Large Dependence High Gray Level Emphasis | -0.32663 | -0.597 ± 0.951 | 0.152 ± 0.765 | <0.05* |
| F6 | Delta Radiomics | Wavelet-HHH | glszm | Gray Level NonUniformity | -0.43211 | -0.379 ± 0.674 | 0.096 ± 0.497 | <0.05* |
| F7 | 1st Radiomics | original | glcm | MCC | 0.51466 | 0.146 ± 0.812 | -0.037 ± 1.054 | 0.486 |
| F8 | Delta Radiomics | Wavelet-LLH | firstorder | Median | 0.69508 | 0.558 ± 1.214 | -0.142 ± 1.086 | 0.064 |
| F9 | Delta Radiomics | Wavelet-HLL | firstorder | Kurtosis | 1.09794 | 0.254 ± 0.831 | -0.065 ± 0.326 | 0.181 |

*** Significant difference is identified based on the Mann-Whitney U test.**

### 3.4 Visualization of Model Discrimination and Multiomics Feature Interpretation

The probability scores obtained using the method of Platt scaling are presented in **Fig. 5A**, which shows the output probability scores of all lesions in different models, demonstrating the varying degrees of discrimination among models. Higher values for individuals in *Group A* and lower values for *Group B* indicate better separability between the two classes. The mean probability scores for *Group A* versus *Group B* are as follows (in the order displayed from top to bottom in **Fig. 5A**): 0.322 ± 0.070 vs. 0.187 ± 0.090 for the 1st radiomics model, 0.510 ± 0.167 vs. 0.145 ± 0.132 for the 2nd radiomics model, 0.614 ± 0.167 vs. 0.129 ± 0.098 for the delta radiomics model, 0.350 ± 0.255 vs. 0.176 ± 0.056 for the 1st dosiomics model, 0.356 ± 0.132 vs. 0.170 ± 0.101 for the 2nd dosiomics model, 0.565 ± 0.204 vs. 0.103 ± 0.093 for the delta dosiomics model, and 0.720 ± 0.196 vs. 0.073 ± 0.053 for the EFS model. Among all the models, the EFS model demonstrates the most effective separation between the two groups, with minimal overlap in probability scores. **Fig. 5B** visualizes the feature space using UMAP dimensionality reduction. The nine features from the EFS model are projected into a three-dimensional space to show the separation between lesions in *Group A* and *Group B*. An SVM hyperplane serves as the decision boundary between two groups. To facilitate differentiation, lesions in *Group A* are marked in red, while lesions in *Group B* are marked in blue.

**Figures 5C** and **5D** display results for two lesions, featuring MRI and dose maps alongside wavelet-transformed images to aid in understanding multiomics feature extraction. **Table 4** outlines the 9 features of the EFS model, primarily derived from radiomics features post-wavelet transformation (3 features directly from the original images). The 3D discrete wavelet transform decomposes volumetric images into multi-resolution components, capturing both fine and coarse details, with "L" representing low-pass and "H" representing high-pass filtering.[30] Features F2 and F9, with the highest weight coefficients within the SVM model (F2: -1.24, F9: 1.10), are elaborated below.

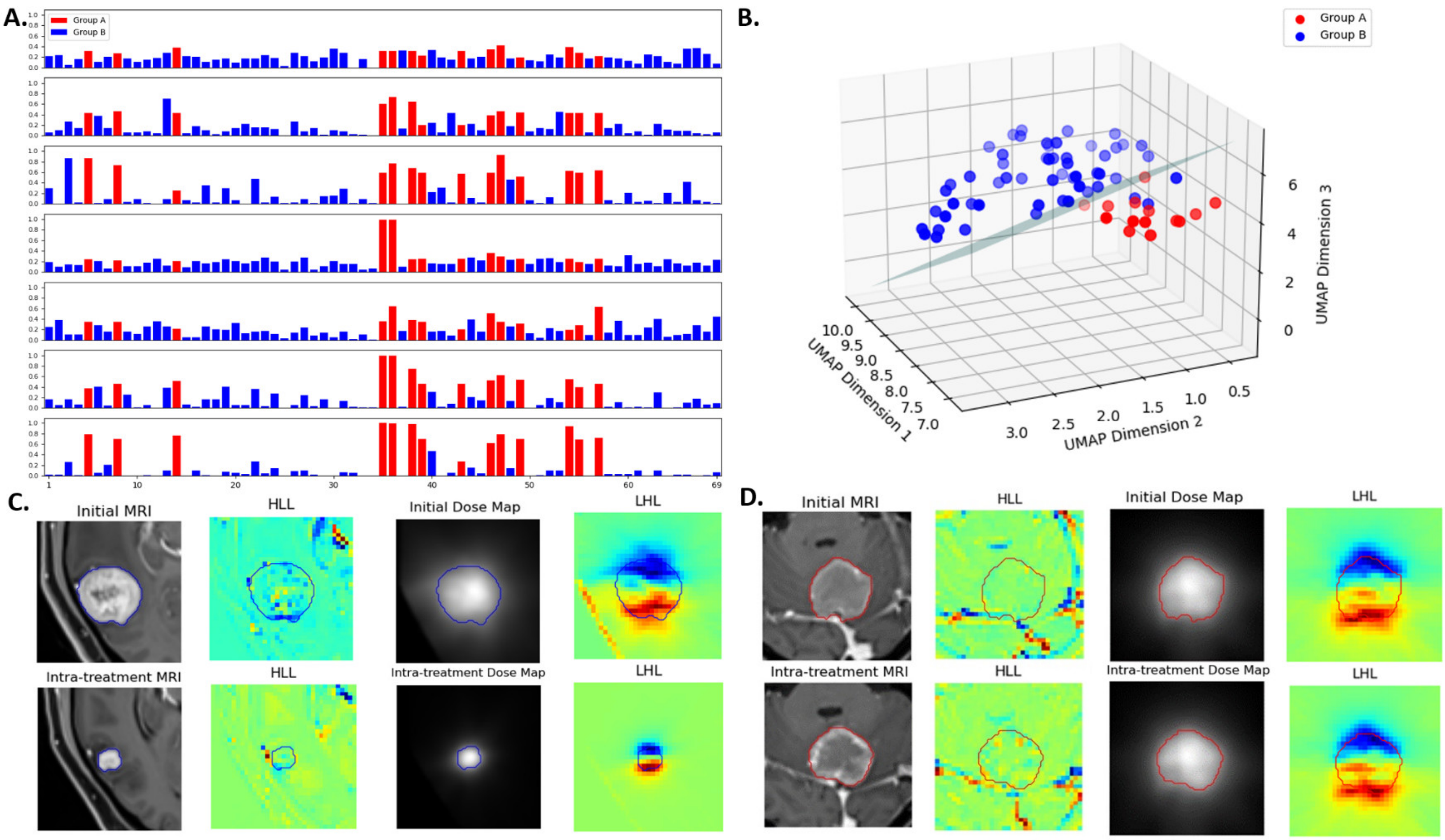

**Figure 5.** (Fig. 5A) Probability scores for lesions obtained across different models, demonstrating varying degrees of discrimination. The horizontal axis represents the lesion index, and the vertical axis represents the probability score of a lesion under different models. A score closer to 1 indicates a higher likelihood of non-decreased GTV at follow-up compared to initial GTV after PULSAR treatment, while a score closer to 0 indicates a higher likelihood of a decreased GTV. The models, from top to bottom, are: 1st radiomics, 2nd radiomics, delta radiomics, 1st dosiomics, 2nd dosiomics, delta dosiomics, and the EFS model. The EFS model shows the most effective separation between the two groups. (Fig. 5B) UMAP visualization of the EFS model's nine features, projected into a three-dimensional space to illustrate the separation between Group A and Group B lesions. An SVM hyperplane serves as the decision boundary between the groups. Non-decreased lesions (Group A) are marked in red, while decreased lesions (Group B) are marked in blue. (Fig. 5C) A lesion (contour in blue) with decreased GTV at follow-up compared to the initial. (Fig. 5D) A lesion (contour in red) with non-decreased GTV at follow-up compared to the initial. Panels include initial MRI, intra-treatment MRI, initial dose map, intra-treatment dose map, and wavelet-transformed images (HLL, LHL) for elucidating multiomics feature extraction.

Feature F2, known as Low Gray Level Zone Emphasis (LGLZE), quantifies the prevalence of low gray-level size zones. It reflects the spatial uniformity in dose maps. The LGLZE values are 0.35 for decreased GTV (**Fig. 5C**) and 0.61 for non-decreased GTV (**Fig. 5D**), respectively. Lower LGLZE values suggest fewer low-dose regions within the tumor, potentially correlating with a favorable treatment response.

Feature F9, derived from Kurtosis calculated in the delta mode based on radiomics, provides insights into evolving tumor heterogeneity during treatment. The lesions in **Figs. 5C** and **5D** exhibit the delta Kurtosis values of 0.21 and 0.36, respectively. The lesion with a smaller delta Kurtosis value shows a reduction in volume post-treatment, while the lesion with a larger delta Kurtosis value exhibits the opposite trend. This tends to suggest that the change in tumor heterogeneity after the first treatment is tied to treatment outcome.

**4. Discussion**

Our study focuses on the treatment of BM within the PULSR framework, utilizing multiomics modeling to predict treatment outcomes and assist with potential decision-making. By integrating multiomics with SVM classification, we aim to achieve a more accurate prediction of treatment response at an earlier stage (e.g., intra-treatment). Notably, our EFS model integrates crucial features from pre-treatment radiomics, pre-treatment dosiomics, intra-treatment radiomics, and delta-radiomics, surpassing six individual models.

A unique feature of PULSAR is the availability of intra-treatment MRI images, which allows for the effective use of delta-omics. This capability led to the development of the EFS model, which demonstrated superior performance by integrating key features from pre-treatment radiomics, pre-treatment dosiomics, intra-treatment radiomics, and delta-radiomics, outperforming traditional radiomics models that choose features solely based on pre-treatment information. This observation is consistent with previous findings.[22-24] In constructing the EFS model, we considered the potential contributions of various feature sets, including the 2nd dosiomics and delta dosiomics features. However, we ultimately chose to exclude these features based on two practical considerations. Firstly, from a decision-making standpoint, the delta dosiomics (and 2nd dosiomics) data would not yet be accessible before a physician makes a new dose prescription and a new treatment plan. Secondly, the classification performance is comparable between delta dosiomics and delta radiomics, suggesting that the added benefits of the delta dosiomics (and 2nd dosiomics) are limited. This raises two relevant questions: 1) Why does the inclusion of delta radiomics features improve performance? 2) What additional information does dosiomics generate?

For the first question, we believe that delta radiomics features are able to capture dynamic biological changes or the evolution of anatomical features within GTV. Inclusion of the temporal change yields more accurate prediction, as reflected by the significant improvement of both delta radiomics (and 2nd radiomics) relative to the 1st radiomics. Put simply, to assess treatment response effectively, it's best to conduct evaluations after the treatment has been administered. Although Gao et al. did not capture the dynamic changes during treatment like our study, they explored radiomics features from longitudinal diffusion-weighted imaging (DWI) to predict post-treatment tumor necrosis. By comparing the prediction performance using different time points, they found that information from a single or multiple time points alone was insufficient to reflect the tumor's response to treatment. Notably, delta features calculated by pre- and post-treatment were more predictive than static features too. This finding emphasizes the importance of including feature changes to capture the tumor's response to radiation.[24] Furthermore, Wang et al. investigated the prognostic value of delta-radiomics features in predicting treatment response and clinical outcomes, such as overall survival (OS) and disease-free survival (DFS). Their study showed that delta-radiomics features outperformed single timepoint radiomics models or clinical models in predicting patient prognosis. These findings underscore the importance of incorporating temporal changes in radiomics analysis to capture the dynamic nature of tumor response to treatment.[23] Essentially, assessing treatment response is more effective when evaluations are conducted post-treatment. For the second question, dosiomics features extracted from 3D dose distributions offer advantages over traditional dose-volume histogram (DVH) analyses.[20,31] By offering a comprehensive pixel-wise representation of radiation dose, dosiomics captures patient-specific dose heterogeneity within the BM and surrounding tissues.[32] Our results indicate that delta dosiomics outperforms second and first dosiomics, suggesting that changes in pre- and post-treatment dose maps, particularly in dose fall-off regions, provide valuable classification information. However, the features from delta radiomics and delta dosiomics may overlap, allowing either to be used for classification with comparable accuracy.

In our study, significant effort was dedicated to feature selection and minimizing redundancy. To ensure a fair comparison among different models while adhering to the guideline of maintaining a feature-to-observer ratio of at least 4:1, we consistently selected the top 9 features for each model.[33] The initial raw feature set contained 851 features, presenting a challenge that was addressed using the MLF strategy detailed in section 2.3. Traditional feature selection methods often utilize a two-sample t-test to identify features with the smallest p-values or employ multi-tiered selection methods such as Lasso regression and Welch's t-test.[34,35] In our analysis, **Table 4** demonstrates that 6 of the 9 selected features have significant p-values below 0.05 using the Mann-Whitney U test. The remaining 3 features, while not statistically significant individually, exhibit positive SVM coefficients (F7, F8, and F9). This suggests that these features, despite their high p-values, contribute significantly to the model's classification performance when considered in conjunction with other features, underscoring the importance of feature correlation.

In the EFS model, we identified the top 9 features, as detailed in section 3.4. While features like tumor size or shape (F1) are straightforward, others derived from texture matrices and wavelet filtering in **Table 4** are more complex to interpret—a common issue in the field. These features may correlate with tissue heterogeneity or dosiomics gradients, with the wavelet transform aiding in multi-resolution decomposition and noise suppression.[30,36] In our next phase of study, we aim to improve interpretability by establishing correlations between the features in **Table 4** and other pathological and clinical features. In particular, integrating the delta change of biomarkers that indicate the response to chemotherapy or immunotherapy into the prediction model may help leverage their synergistic effects, highlighting another unique feature of PULSAR treatment.[11,37,38] Additionally, while our current focus is primarily on features extracted from within GTV, valuable insights may also lie in features from the peritumoral region. Gradient maps in the lesion's periphery, for example, could serve as promising indicators of treatment response.[39,40] Furthermore, exploring CNN-based auto-encoders shows promise for feature extraction due to their translation equivariance property, potentially offering more interpretable results compared to multiomics approaches.

Three limitations in our current study warrant further investigation. Firstly, the patient cohort size was small and retrospective due to the early phase of the PULSAR trial at UTSW, resulting in a limited number of cases meeting the relatively stringent recruitment criteria. Efforts are currently underway to expand the cohort size, aiming to enhance the accuracy and robustness of the prediction. Additionally, the imbalanced nature of the study cohort presents another limitation, compounded by the absence of a completely independent dataset for validation. Finally, tumor size was selected as the prediction outcome due to evidence suggesting that changes in tumor volume are valid indicators of tumor regression or progression.[13-15,41] For instance, a reduction in tumor volume exceeding 25% is significantly associated with a higher complete response rate.[42] However, further examination is needed to determine whether volume measured at follow-up (e.g., 2-3 months) correlates with local control and toxicity in BM treatment.

## 5. Conclusion

This study presents a data-driven multiomics approach that uses radiomics and dosiomics to predict tumor volume changes in BM patients undergoing PULSAR treatment. The EFS model exhibits superior performance compared to six individual models, highlighting the significance of incorporating both pretreatment and intra-treatment data in a delta mode. This utilization of multiomics and SVM classification for intra-treatment decision support in PULSAR holds promise for optimizing BM management, potentially mitigating risks associated with under- or over-treatment.

## 6. Acknowledgments

We would like to express our sincere gratitude to Dr. Weiguo Lu, Dr. Chien-Yi Liao, and Mr. Yan Dai for their valuable insights and suggestions during the early stages of this research.

## 7. Conflict of Interest Statement

The authors have no relevant conflicts of interest to disclose.

## 9. Figure Legend

**Figure 1.** Comparison of workflow between (A) fractionated stereotactic radiotherapy (fSRT) and (B) personalized ultra-fractionated stereotactic adaptive radiotherapy (PULSAR). PULSAR includes an intra-treatment MRI assessment to evaluate the change of GTV (increased, unchanged, or decreased), enabling more personalized treatment and timely adjustment.

**Figure 2.** Six lesions illustrate the temporal evolution of GTV at various treatment stages. In Group A (lesions with red contours), three lesions exhibit non-decreased GTV at follow-up compared to the initial, but they display different GTV changes at the intra-treatment time point, with intra-treatment assessments of (A) decreased, (B) unchanged, and (C) increased GTV. In contrast, Group B (lesions with blue contours) depicts three lesions with a decreased GTV at follow-up compared to the initial, with intra-treatment variations categorized as (D) decreased, (E) unchanged, and (F) increased GTV.

**Figure 3.** Lesion volumetric changes are examined at three time points (pre-treatment, intra-treatment, and follow-up) for two subsets: lesions smaller than 4000 $mm^3$ (Fig. 3A) and those larger than 4000 $mm^3$ (Fig. 3B). The volumetric dynamics of all 69 lesions are shown with the initial GTV normalized to 1 for each lesion. The line graphs on the left display the relative GTV changes for each lesion during PULSAR. The color intensity of each lesion's trajectory line reflects the relative change at follow-up compared to the initial GTV, with lighter colors representing more significant reductions after PULSAR treatment. Red diamonds indicate lesions with increased GTV at follow-up, while blue triangles represent lesions with decreased GTV at follow-up. The accompanying heatmaps on the right provide a detailed quantitative representation of these changes, with color intensity reflecting the relative increase or decrease in tumor volume.

**Figure 4.** (Fig. 4A) Receiver Operating Characteristic (ROC) curves for six individual models. For each model, the plot shows the aggregated training (solid line) and test (dashed line) ROC curves generated from 50 iterations of 5-fold stratified cross-validation. The mean Area Under the Curve (AUC) with 95% confidence interval is reported for each model, respectively. (Fig. 4B) Performance evaluation of the Ensemble Feature Selection (EFS) model. The left panel displays the coefficient values of the 9 selected features in the EFS model. The middle panel presents the correlation heatmap of these 9 EFS features, demonstrating minimal inter-feature correlation. The right panel shows the ROC curve of the EFS model similar to those in (A).

**Figure 5.** (Fig. 5A) Probability scores for lesions obtained across different models, demonstrating varying degrees of discrimination. The horizontal axis represents the lesion index, and the vertical axis represents the probability score of a lesion under different models. A score closer to 1 indicates a higher likelihood of non-decreased GTV at follow-up compared to initial GTV after PULSAR treatment, while a score closer to 0 indicates a higher likelihood of a decreased GTV. The models, from top to bottom, are: 1st radiomics, 2nd radiomics, delta radiomics, 1st dosiomics, 2nd dosiomics, delta dosiomics, and the EFS model. The EFS model shows the most effective separation between the two groups. (Fig. 5B) UMAP visualization of the EFS model's nine features, projected into a three-dimensional space to illustrate the separation between Group A and Group B lesions. An SVM hyperplane serves as the decision boundary between the groups. Non-decreased lesions (Group A) are marked in red, while decreased lesions (Group B) are marked in blue. (Fig. 5C) A lesion (contour in blue) with decreased GTV at follow-up compared to the initial. (Fig. 5D) A lesion (contour in red) with non-decreased GTV at follow-up compared to the initial. Panels include initial MRI, intra-treatment MRI, initial dose map, intra-treatment dose map, and wavelet-transformed images (HLL, LHL) for elucidating multiomics feature extraction.